# Training deep neural networks for the inverse design of nanophotonic structures

Dianjing Liu, Yixuan Tan, Erfan Khoram, and Zongfu Yu*

Department of Electrical and Computer Engineering, University of Wisconsin, Madison, Wisconsin 53706, U.S.A

**Abstract:** Data inconsistency leads to a slow training process when deep neural networks are used for the inverse design of photonic devices, an issue that arises from the fundamental property of non-uniqueness in all inverse scattering problems. Here we show that by combining forward modeling and inverse design in a tandem architecture, one can overcome this fundamental issue, allowing deep neural networks to be effectively trained by data sets that contain non-unique electromagnetic scattering instances. This paves the way for using deep neural networks to design complex photonic structures that requires large training sets.

**Keywords:** Nanophotonics; Inverse scattering; Neural networks

*zyu54@wisc.edu*

Today's nanophotonic devices increasingly rely on complex nanostructures to realize sophisticated functionalities. As structural complexity grows, the design processes become more challenging. Conventional design approaches are based on optimization. One typically starts with a random design and computes its response using electromagnetic simulations. The result is compared to the target response, and a structural change is calculated to update the design. This process is performed iteratively. Notable examples include evolutionary algorithms[1], level set methods[2], adjoint methods[3], and the optimization of specific geometric parameters[4–6]. It often takes hundreds or even thousands of simulations before a reasonable design can be found. Since each simulation is computationally expensive, these methods become prohibitively slow as device size and complexity grow.

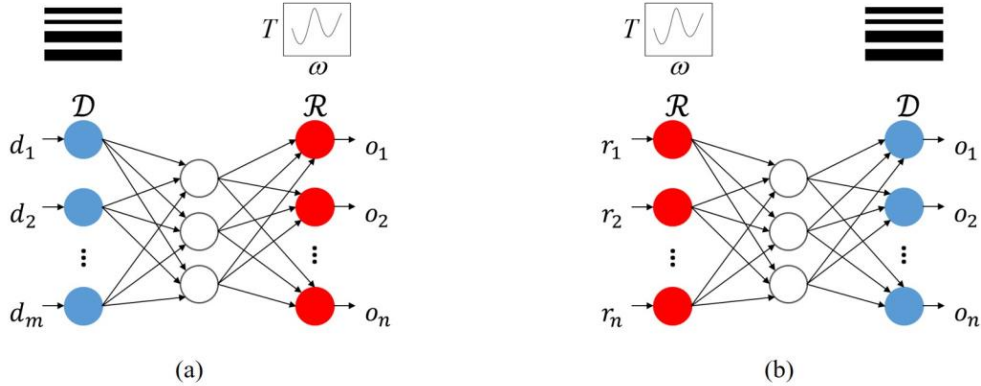

FIG. 1. (a) A forward modeling neural network with one hidden layer. The neural network takes device designs $\mathcal{D}$ as inputs and outputs corresponding responses $\mathcal{R}$. (b) An inverse network with one hidden layer. It takes device responses $\mathcal{R}$ as the input and outputs the designs $\mathcal{D}$.

In contrast to the optimization approach, data-driven approaches based on machine learning are rapidly emerging, where artificial neural networks (NNs)[7–12] are trained to assist in the design process[13]. NNs can be used in two different ways. The first method, as shown in Fig. 1(a), takes the input of the structural parameter $\mathcal{D}$ (such as the geometrical shape of a nanostructure) and predicts the electromagnetic response $\mathcal{R}$ of the device (such as transmission spectra or differential scattering cross-section). These NNs are used to replace the computationally expensive EM simulations in the optimization loop, greatly reducing design time[13,14]. We refer to these NNs as forward-modeling networks because they compute EM response from the structure. In contrast, the second type of NNs, as shown in Fig. 1(b), take the EM response as the input and directly output the structure. These are referred to as inverse-design networks. These NNs can accomplish the design goal in a fraction of second without needing any iterative optimization. For both forward-modeling and inverse-design networks, one needs a large amount of training instances ($\mathcal{R}_i$, $\mathcal{D}_i$) to train the networks before they can perform the intended function. Creating these training instances involves electromagnetic simulations and can require significant amounts of computational resources. However, this is a one-time cost. In contrast, conventional optimization requires the same large amount of simulations for each design. This is the key advantage of the data-driven method: simulations are invested in to build the design tool, while they are constantly consumed in conventional optimization methods.

Training for forward modeling can be done in a standard neural network. On the other hand, there has been one significant challenge in training deep NNs for inverse design. This arises from a fundamental property of the inverse scattering problem: the same EM response $\mathcal{R}$ can be created by many different designs $\mathcal{D}$. This non-unique response-to-design mapping creates conflicting training instances, such as ($\mathcal{R}$, $\mathcal{D}^A$) and ($\mathcal{R}$, $\mathcal{D}^B$). When such conflicting instances with the same input but different output labels exist in the training data set, the neural network would be hard to converge.

Early work[15] tried to solve this problem by dividing the training data set into distinct groups, so that within each group there is unique design $\mathcal{D}$ corresponding to each response $\mathcal{R}$. This method demonstrated limited success on small training sets. As we will show later, eliminating the apparent conflicting instances does not fundamentally address the issue of non-unique mapping, and thus is generally ineffective. Here we propose a tandem network structure to solve the issue. By cascading an inverse-design network with a forward-modeling network, the tandem network can be trained effectively.

First, we use a specific example to illustrate the difficulty in training deep NNs for inverse design. As shown in Fig. 2(a), we consider a thin film consisting of alternating layers of $SiO_2$ and $Si_3N_4$. The goal is for this multi-layer film to generate a target transmission spectrum; the design space is the thickness of each layer. The structure can be represented by an array $\mathcal{D} = [d_1, d_2, ..., d_m]$, with $d_i$ being the thickness of the i$^{th}$ layer. The transmission spectrum is discretized by $n$ points, and represented by an array $\mathcal{R} = [r_1, r_2, ..., r_n]$. We set the maximum thickness of each layer to be $a$. The spectral range of interest is $0.15c/a \leq f \leq 0.25c/a$.

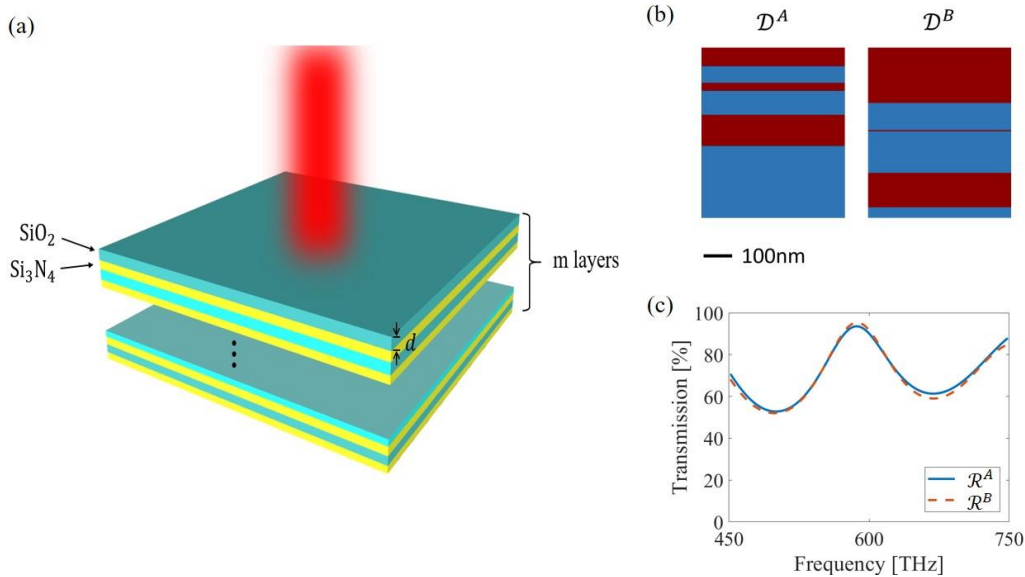

FIG. 2. (a) A thin film composed of m layers of $SiO_2$ and $Si_3N_4$. The design parameters of the thin film are the thicknesses of the layers $d_i$ (i=1, 2, ..., m) and the device response is the transmission spectrum. The forward neural network takes $\mathcal{D}=[d_1, d_2, ..., d_m]$ as inputs and discretized transmission spectrum $\mathcal{R}$ as output. (b)(c) Example of two 6-layer thin film designs with very similar transmission spectra.

We use full-wave EM simulations to generate training instances, where we solve the transmission spectrum $\mathcal{R}$ for a randomly generated structure $\mathcal{D}$. The number of instances ($\mathcal{R}, \mathcal{D}$) typically ranges from tens to hundreds of thousands. In practice, the training data set may not include instances with identical response. However, as long as there are instances with distinct structures and almost the same transmission spectra, the training of the neural network would be hard to converge. For example, the two instances from the training data have structures $\mathcal{D}^A$ and $\mathcal{D}^B$, as shown in Fig. 2(b). These two films turn out to have almost identical transmission spectra $\mathcal{R}^A \approx \mathcal{R}^B$, as shown in Fig. 2(c). When we have both instances ($\mathcal{R}^A, \mathcal{D}^A$) and ($\mathcal{R}^B, \mathcal{D}^B$) in the training set, the training will struggle to converge, as the two instances give the network completely different answers $\mathcal{D}^A$ and $\mathcal{D}^B$ for a slight input change from $\mathcal{R}^A$ to $\mathcal{R}^B$.

We can consider a specific network to examine the training process. Training is done by minimizing a cost function, for example $E = \frac{1}{2}\sum_i (d_i - o_i)^2$, where $o_i$ is the layer thickness designed by the neural network given the input $\mathcal{R}$, and $d_i$ is the ground truth of the layer thickness. The cost function measures the distance between the prediction of the network $O$ and the ground truth $\mathcal{D}$ used in simulation. We use a fully connected network of four layers. Its architecture is denoted as $200 - 500 - 200 - 20$, with the figures indicating the number of units in each layer. The network has an input layer of 200 units (n=200), which matches the number of discretization points of the transmission spectrum. The output layer of 20 units (m=20) indicates the layer thickness of a 20-layer film. It has two hidden layers with 500 and 200 units respectively. The training set includes 500,000 instances, while another different 50,000 instances are left as the test set. The learning curve is shown in Fig. 3(a) (blue line). The cost function barely drops even after 15,000 epochs of training, indicating the network's poor performance in designing the thin film structure for the input transmission spectrum. Increasing the size of the inverse network or tuning hyperparameters such as the learning rate does not improve its performance either. As shown in Fig. 3(b), the design produced by this NN turns out to be far off from the target spectrum. This observation is consistent with previous studies[15,16].

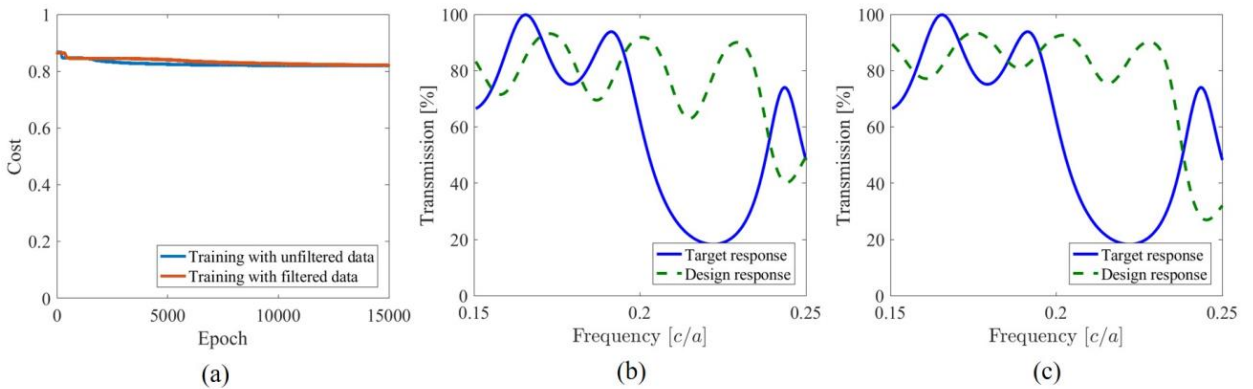

FIG. 3. (a) Learning curve of the inverse network. The blue line is directly trained by 500,000 instances. The red line is trained by filtered the training data (26.1% instances remain). The learning rate is initially $10^{-4}$ and decreases by half every 3,000 epochs. Hyperparameters are chosen by grid search and are the same for all trainings. (b) Test example of the inverse network trained by an unfiltered training set. (c) Test example of the inverse network trained by a filtered training set.

One might be tempted to resolve this issue by eliminating the non-unique instance in the training data set. This can be done, for example, by defining a distance between two transmission spectra $\mathcal{R}^{(1)}$ and $\mathcal{R}^{(2)}$ as $\frac{1}{2}\sum_i (r_i^{(1)} - r_i^{(2)})^2$. We can then remove one of the two training instances when their distance falls below a threshold. This filtering method was used[15] with limited success in a small data set. Here, we applied the same approach, and as can be seen by the red line in Fig. 3(a), filtering the training instance barely improves the training (the test example is shown in Fig. 3(c)). Even without apparently conflicting instances, there are still implicitly conflicting instances that cannot be easily eliminated.

To understand the origin of these implicit conflicting instances, let us assume that there is an ideal training set without any explicit or implicit conflicting instances $S = \{<\mathcal{R}_1, \mathcal{D}_1>, <\mathcal{R}_2, \mathcal{D}_2>, ..., <\mathcal{R}_n, \mathcal{D}_n>\}$ that allows the training to converge successfully. This training set can be easily contaminated to include conflicting instances. To show such an instance, we can first train a network based on the ideal training set $S$. Then we take an arbitrary $\mathcal{R}_0$ that is different from all $\mathcal{R}$ in the training set $S$, which outputs $\mathcal{D}_0$. The instance $<\mathcal{R}_0, \mathcal{D}_0>$ is consistent with training set $S$. In electromagnetic scattering, there are often other solutions, for example $\mathcal{D}_0^r$ with $\mathcal{D}_0^r \neq \mathcal{D}_0$, that also generate the same response $\mathcal{R}_0$. Now if a training set $S'$ contains $<\mathcal{R}_0, \mathcal{D}_0^r>$, i.e. $S'=\{S, <\mathcal{R}_0, \mathcal{D}_0^r>\}$, there is no apparent one-to-many mapping issue. However, when training with $S'$, the instance $<\mathcal{R}_0, \mathcal{D}_0^r>$ would pull the network away from the prediction $\mathcal{D}_0$, which would be the right prediction if it were trained by $S$. The presence of this new instance makes the convergence slow, if it ever converges at all. And the presence of such inconsistency is difficult to detect and cannot be eliminated by the filtering method (see *Supporting Information* for further discussion).

Now we introduce our method to overcome the above issue. We propose a tandem architecture consisting of two neural networks as shown in Fig. 4. The first is the same as the traditional network for the inverse design, and the second part is a forward network trained to predict the response of a design. When using the tandem network for inverse design, a desired response $\mathcal{R}$ is taken as the input. The output by the intermediate layer $M$ (shown in Fig. 4) is the designed structure. The output of the tandem network is the response calculated from the designed structure. The forward modeling network is pre-trained first. Then, the weights in the pre-trained forward modeling network are fixed and the weights in the inverse network are trained to reduce the cost function defined as the error between the predicted response and the target response. This network structure overcomes the issue of non-uniqueness in the inverse scattering of electromagnetic waves because the design by the neural network is not required to be the same as the real design in training samples. Instead, the cost function would be low as long as the generated design and the real design have similar response.

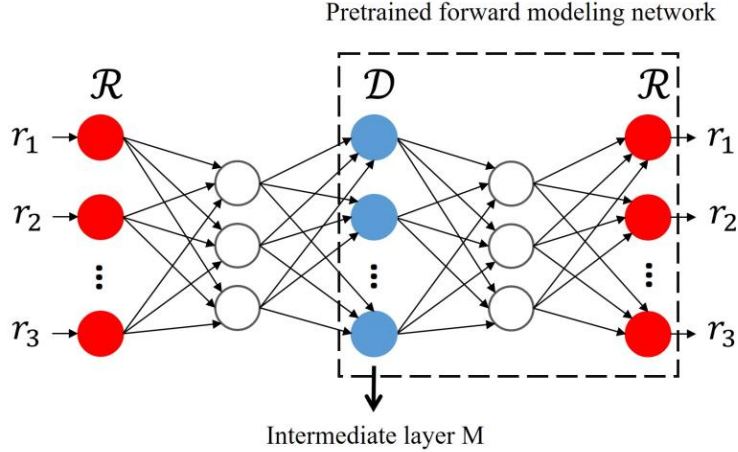

FIG. 4. A tandem network is composed of an inverse design network connected to a forward modeling network. The forward modeling network is trained in advance. In the training process, weights in the pre-trained forward modeling network are fixed and the weights in the inverse network are adjusted to reduce the cost (i.e., error between the predicted response and the target response). Outputs by the intermediate layer M (labeled in blue) are designs $\mathcal{D}$.

In training, we first separate the second part of the network, i.e., the forward-modeling network, and independently train this network with training instances obtained from full-wave electromagnetic simulations. The input of the forward network is the design $\mathcal{D}$, and the output is the response $\mathcal{R}$. As there is always a unique response $\mathcal{R}$ for every design $\mathcal{D}$, the training is easy to converge (see *Supporting Information* for detailed implementation).

With successful training of the forward-modeling network, we now connect it to an inverse-design network to form a tandem neural network (as shown in Fig. 4). The inverse network architecture is set to have four layers with each layer having $200 - 500 - 200 - 20$ units. The spectrum $\mathcal{R} = [r_1, r_2, ..., r_{200}]$ is taken as the input of the tandem network. A design $\mathcal{D}$ is calculated as the intermediate layer, which is then fed into the forward modeling part to calculate the corresponding spectrum $[o_1, o_2, ..., o_{200}]$. The training is done by minimizing the cost function of the tandem network defined as $E = \frac{1}{2}\sum_i (r_i - o_i)^2$. As shown by the learning curve in Fig. 5(a), the rapidly decreasing cost of test instances shows that training is highly effective. Indeed, the structures designed by the tandem network create the desired transmission spectra with much better fidelity, as shown in Fig. 5(b) and (c).

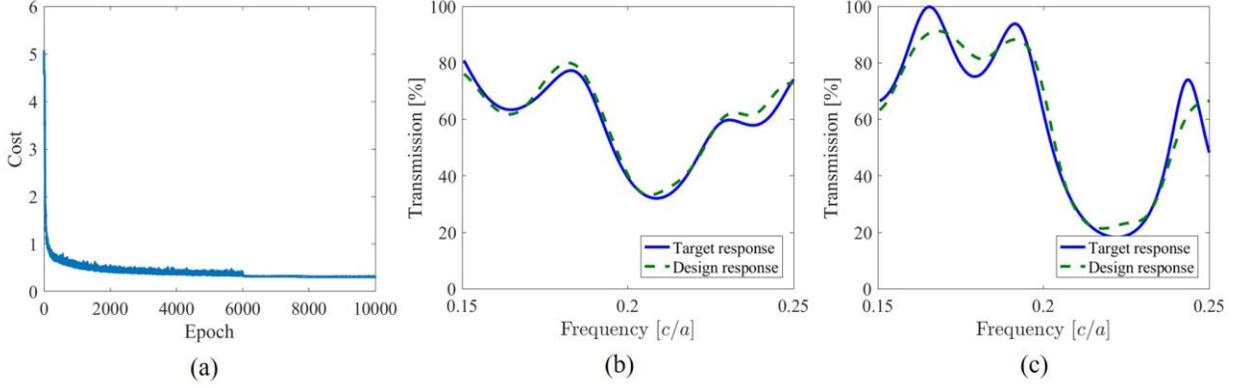

FIG. 5. (a) Learning curve of the tandem neural network. (b) (c) Example test results for the tandem network method.

Here we show a specific example of designing the structure of 16-layer $SiO_2$ and $Si_3N_4$ thin film for certain target transmission spectra. The maximum thickness of each layer is set to be 150 nm. The response is the transmission spectrum within the range of 300 to 750 THz, corresponding to wavelength λ of 400 to 1000 nm. The target transmission spectra are set to be of a Gaussian shape:

$$t(f) = 1 - 0.8 \exp\left[-\frac{(f-f_0)^2}{2\sigma^2}\right], \qquad (1)$$

Here $f_0 = 525$ THz and $\sigma$ is set to be 75, 37.5 and 18.75 THz for three cases. The corresponding spectra are shown in Fig. 6 (blue lines). For the three target spectra, the tandem network designs structures as follows:

$o^{(1)}$ = [79, 72, 100, 107, 68, 20, 8, 53, 101, 91, 78, 61, 70, 104, 108, 12],

$o^{(2)}$ = [118, 106, 114, 100, 36, 38, 16, 48, 81, 122, 26, 92, 48, 122, 127, 4],

$o^{(3)}$ = [111, 111, 132, 101, 27, 51, 26, 33, 59, 141, 8, 104, 16, 128, 137, 4].

(Unit: nm)

The spectra of the designed structures are shown in Fig. 6 (green dashed lines), and reasonably satisfy the design goal. It only takes a fraction of a second for the neural network to compute a design. We expect the performance of the design can be further improved with more training instances.

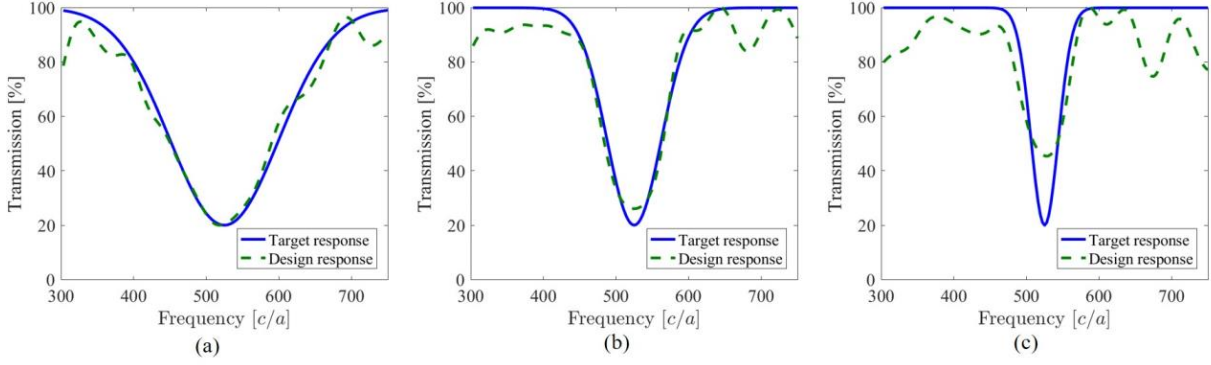

FIG. 6. Example design by the tandem neural network. The blue lines are Gaussian shaped target spectra, and the green lines are spectra of tandem network designs.

Finally, we demonstrate an example of designing 2D structures to modulate transmission phase delay independently at three wavelengths: R (470 nm), G (540 nm), B (667.5 nm). In order to make the problem more trackable, we parameterize the structures to reduce data dimension. The designed units are composed of 3 layers of Si and $SiO_2$ as shown in Fig. 7. Within each layer, part of Si or $SiO_2$ is removed to form a rectangular slot. The design parameters are thicknesses of the 3 layers $d_i$ ($i = 1,2,3$), the location $x_i$ and width $w_i$ of the vacuum slot in the $i$th layer ($i = 1,2,3$). The thickness $d_i$ of each layer ranges from 150 to 450nm. The unit width is 400nm. This meta unit can be used in a metasurface to create three-color holograms[17,18].

We use Rigorous Coupled Wave Analysis (RCWA) method[19] to simulate phase delay of the randomly generated structures. The incident light is $s$ polarization and is along $+y$ direction. In the $z$–direction the material is homogeneous and in the $x$–direction periodic boundary condition is applied. The training data set includes 750,000 instances and test data set includes 5,000 instances. Training details are included in *Supporting Information*.

The phase delay of the designed structure has an average error of 16.0°. We randomly pick three cases and list target and design responses in Table 1. The designed structures are shown in Fig. 7 and corresponding parameters are in Table 2.

Table 1: Example design result by tandem network. $\phi_R$, $\phi_G$ and $\phi_B$ are phase delay at R (470nm), G (540nm) and B (667.5nm) wavelengths.

| Case | Target response | | | Design response | | |
|---|---|---|---|---|---|---|
| | $\phi_R$ | $\phi_G$ | $\phi_B$ | $\phi_R$ | $\phi_G$ | $\phi_B$ |
| a | 163.0° | −72.6° | −4.2° | 153.9° | −88.6° | −3.0° |
| b | −93.7° | 119.1° | −157.3° | −97.9° | 123.6° | −170.6° |
| c | −147.8° | 78.8° | 169.8° | −137.3° | 75.8° | 154.4° |

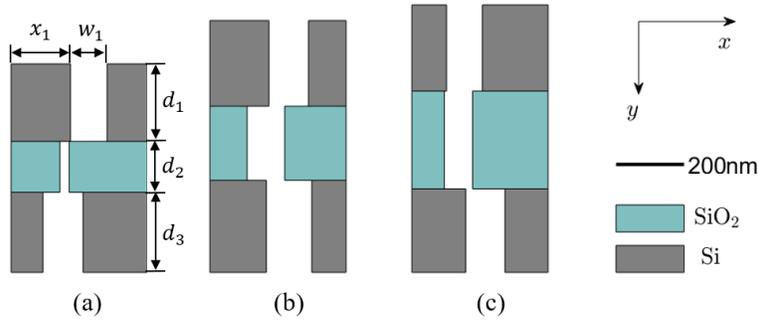

FIG.7. Designed structures of cases (a) (b) and (c) in Table 1.

Table 2: Design parameters of structures in Fig. 7. (Unit: nm)

| Case | $d_1$ | $x_1$ | $w_1$ | $d_2$ | $x_2$ | $w_2$ | $d_3$ | $x_3$ | $w_3$ |
|------|-------|-------|-------|-------|-------|-------|-------|-------|-------|
| a    | 288   | 175   | 108   | 189   | 143   | 28    | 297   | 94    | 118   |
| b    | 318   | 173   | 116   | 275   | 109   | 110   | 341   | 166   | 133   |
| c    | 319   | 102   | 106   | 363   | 95    | 84    | 309   | 158   | 115   |

In conclusion, we show that using neural networks for the inverse design suffers from the problem of non-uniqueness, a typical issue in the inverse scattering problem. This issue makes it very difficult to train neural networks on a large training data set, which is often needed to model complex photonic structures. Here we demonstrate a tandem architecture that tolerates both explicit and implicit non-unique training instances. It provides a way to train large neural networks for the inverse design of complex photonic structures.

**Acknowledgement**

The authors acknowledge the financial support of DARPA YFA program (YFA17 N66001-17-1- 4049).**Supporting Information:**

I.   Data consistency in inverse design problems.

II.  Training forward neural network.

III. Training neural network to design transmission phase delay of 2D structure.

(1) Gondarenko, A.; Lipson, M. Low Modal Volume Dipole-like Dielectric Slab Resonator. *Opt. Express* **2008**, *16* (22), 17689–17694.

(2) Kao, C. Y.; Osher, S.; Yablonovitch, E. Maximizing Band Gaps in Two-Dimensional Photonic Crystals by Using Level Set Methods. *Appl. Phys. B* **2005**, *81* (2–3), 235–244.
(3) Piggott, A. Y.; Lu, J.; Lagoudakis, K. G.; Petykiewicz, J.; Babinec, T. M.; Vučković, J. Inverse Design and Demonstration of a Compact and Broadband on-Chip Wavelength Demultiplexer. *Nat. Photonics* **2015**, *9* (6), 374–377.
(4) Seliger, P.; Mahvash, M.; Wang, C.; Levi, A. Optimization of Aperiodic Dielectric Structures. *J. Appl. Phys.* **2006**, *100* (3), 034310.
(5) Oskooi, A.; Mutapcic, A.; Noda, S.; Joannopoulos, J. D.; Boyd, S. P.; Johnson, S. G. Robust Optimization of Adiabatic Tapers for Coupling to Slow-Light Photonic-Crystal Waveguides. *Opt. Express* **2012**, *20* (19), 21558–21575.
(6) Shen, B.; Wang, P.; Polson, R.; Menon, R. An Integrated-Nanophotonics Polarization Beamsplitter with 2.4$\times$ 2.4 μm2 Footprint. *Nat. Photonics* **2015**, *9* (6), 378–382.
(7) Rumelhart, D. E.; Hinton, G. E.; Williams, R. J. Learning Representations by Back-Propagating Errors. *Cogn. Model.* **1988**, *5* (3), 1.
(8) Hornik, K.; Stinchcombe, M.; White, H. Multilayer Feedforward Networks Are Universal Approximators. *Neural Netw.* **1989**, *2* (5), 359–366.
(9) Hopfield, J. J. Neural Networks and Physical Systems with Emergent Collective Computational Abilities. In *Spin Glass Theory and Beyond: An Introduction to the Replica Method and Its Applications*; World Scientific, 1987; pp 411–415.
(10) Farhat, N. H.; Psaltis, D.; Prata, A.; Paek, E. Optical Implementation of the Hopfield Model. *Appl. Opt.* **1985**, *24* (10), 1469–1475.
(11) Shen, Y.; Harris, N. C.; Skirlo, S.; Prabhu, M.; Baehr-Jones, T.; Hochberg, M.; Sun, X.; Zhao, S.; Larochelle, H.; Englund, D.; Soljačić, M. Deep Learning with Coherent Nanophotonic Circuits. *Nat. Photonics* **2017**, *11*, 441.
(12) Hermans, M.; Burm, M.; Van Vaerenbergh, T.; Dambre, J.; Bienstman, P. Trainable Hardware for Dynamical Computing Using Error Backpropagation through Physical Media. *Nat. Commun.* **2015**, *6*, 6729.
(13) Peurifoy, J. E.; Shen, Y.; Jing, L.; Cano-Renteria, F.; Yang, Y.; Joannopoulos, J. D.; Tegmark, M.; Soljacic, M. Nanophotonic Inverse Design Using Artificial Neural Network. In *Frontiers in Optics*; Optical Society of America, 2017; p FTh4A–4.
(14) Vai, M. M.; Wu, S.; Li, B.; Prasad, S. Reverse Modeling of Microwave Circuits with Bidirectional Neural Network Models. *IEEE Trans. Microw. Theory Tech.* **1998**, *46* (10), 1492–1494.
(15) Kabir, H.; Wang, Y.; Yu, M.; Zhang, Q.-J. Neural Network Inverse Modeling and Applications to Microwave Filter Design. *IEEE Trans. Microw. Theory Tech.* **2008**, *56* (4), 867–879.
(16) Selleri, S.; Manetti, S.; Pelosi, G. Neural Network Applications in Microwave Device Design. *Int. J. RF Microw. Comput.-Aided Eng.* **2002**, *12* (1), 90–97.
(17) Yu, N.; Capasso, F. Flat Optics with Designer Metasurfaces. *Nat. Mater.* **2014**, *13* (2), 139–150.
(18) Zheng, G.; Mühlenbernd, H.; Kenney, M.; Li, G.; Zentgraf, T.; Zhang, S. Metasurface Holograms Reaching 80% Efficiency. *Nat. Nanotechnol.* **2015**, *10* (4), 308–312.
(19) Liu, V.; Fan, S. S 4: A Free Electromagnetic Solver for Layered Periodic Structures. *Comput. Phys. Commun.* **2012**, *183* (10), 2233–2244.

# I. Data consistency in inverse design problems

The issue of data consistency in training data can be shown with the following example. Let $X$ be an 8×1 real vector and $Y$ be a 4×1 real vector (i.e., $X \in R^8$, $Y \in R^4$), while a nonlinear operator $\hat{O}$ defines a many-to-one mapping from the $X$ space to the $Y$ space:

$$Y = \hat{O}X. \tag{S1}$$

The forward problem, i.e., calculating $Y$ from $X$, is well-defined, and can be solved by training a forward neural network. However, when taking $Y$ as the input and $X$ as the output, the inverse network cannot be trained accurately. The following experiment shows that this is not only caused by non-unique instances in the training data, but also by inconsistency of the data set.

Let $\hat{O}_1^{-1}$ and $\hat{O}_2^{-1}$ be two different operators. For $\forall Y \in R^4$, the two operators satisfy

$$\hat{O}(\hat{O}_1^{-1}Y) = Y. \tag{S2}$$

$$\hat{O}(\hat{O}_2^{-1}Y) = Y. \tag{S3}$$

We generate data set $D_1$ from $\hat{O}_1^{-1}$ so that for each instance $< X_i, Y_i > \in D_1$, $X_i = \hat{O}_1^{-1}Y_i$. In this case, we say the data set $D_1$ is self-consistent, since instances in $D_1$ are sampled from the same mapping $\hat{O}_1^{-1}$. Another self-consistent data set $D_2$ is generated from $\hat{O}_2^{-1}$ in the same way. When $D_1$ and $D_2$ are put together to get a new data set $D_3 = D_1 \cup D_2$, the data set $D_3$ is not self-consistent.

The data set $D_1$, $D_2$, $D_3$ is used to train the inverse network, and the learning curves are shown in Fig. S1. The inverse networks are well trained by $D_1$ and $D_2$. However, the inconsistent data set $D_3$ cannot train an accurate neural network, even though instances are unique in $D_3$ (i.e., all instances have different $Y$ values in $D_3$).

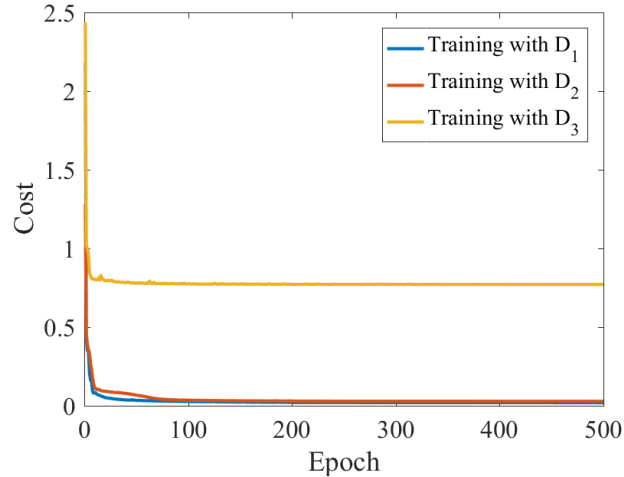

FIG. S1. Learning curve of an inverse network trained by data set $D_1$, $D_2$ and $D_3$. The sets $D_1$ and $D_2$ are self-consistent and can train accurate networks. The set $D_3$ fails to train an accurate network even though instances are unique within $D_3$.

## II. Training forward neural network

In the following, we describe a specific implementation of the forward modelling network training process. To train the forward-modeling network for the multi-layer transmission problem, we experiment with networks having different sizes and depths. Fig. 6(a) compares the learning curves of the networks with different hidden layers. The architectures are as follows.

Architecture 1:  $20 - 500 - 200$

Architecture 2:  $20 - 500 - 200 - 200$

Architecture 3:  $20 - 500 - 200 - 200 - 200$

Architecture 4:  $20 - 500 - 200 - 200 - 200 - 200$

The 20 at the beginning and the 200 at the end are the numbers of input and output units, respectively. As the network becomes deeper, the error decreases, indicating more accurate predictions by the neural network. The network with four hidden layers (i.e., Architecture 4) has error $\approx 0.19$ after 10,000 epochs of training.

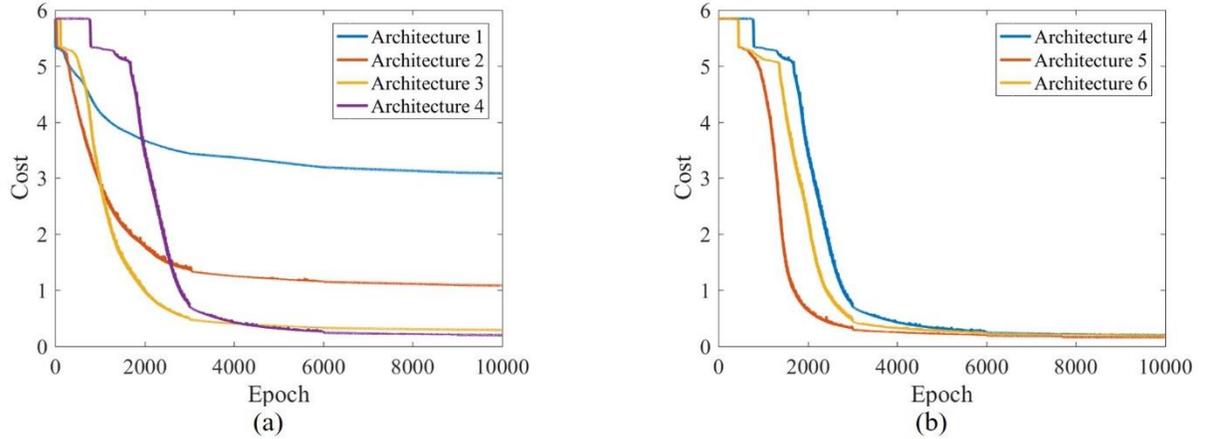

FIG. S2. (a) The learning curve for forward networks with different hidden layers. Architectures 1 to 4 have 1, 2, 3, and 4 hidden layers respectively. (b) The learning curve for forward networks with the same depth but different network sizes.

Fig. S2(b) compares networks with the same depth but different network sizes (number of hidden units in the hidden layers). The architectures are as follows.

Architecture 4:  $20 - 500 - 200 - 200 - 200 - 200,$

Architecture 5:  $20 - 500 - 500 - 200 - 200 - 200,$

Architecture 6:  $20 - 500 - 500 - 500 - 200 - 200.$

The results indicate that larger networks could be trained faster, although as the training goes on, the ultimate performance differs very little.

The network with Architecture 5 has an error ≈0.16 after 12,000 epochs of training. Fig. S3 shows its predictions on three instances randomly chosen from the test set. The ground truth (true transmission spectra) is shown in blue lines for comparison.

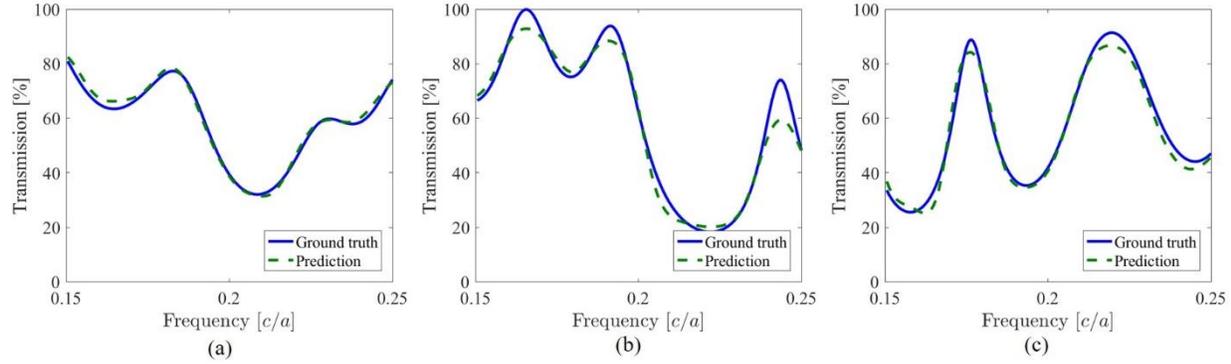

FIG. S3. Example test results of the forward network. The predictions by the network fit well with the ground truth.

## III. Training neural network to design transmission phase delay of 2D structure

When designing 2D structures to modulate transmission phase delay, the forward modeling neural network has 6 hidden layers with each layer having $1024 - 512 - 512 - 256 - 256 - 128$ hidden units. The inverse design network has 2 hidden layers with 512 and 256 hidden units. The learning rate is initially 0.0005 and exponentially decays to $10^{-6}$ at the end of the training. Learning curves of the forward modeling network and the tandem network are shown in Fig. S4.

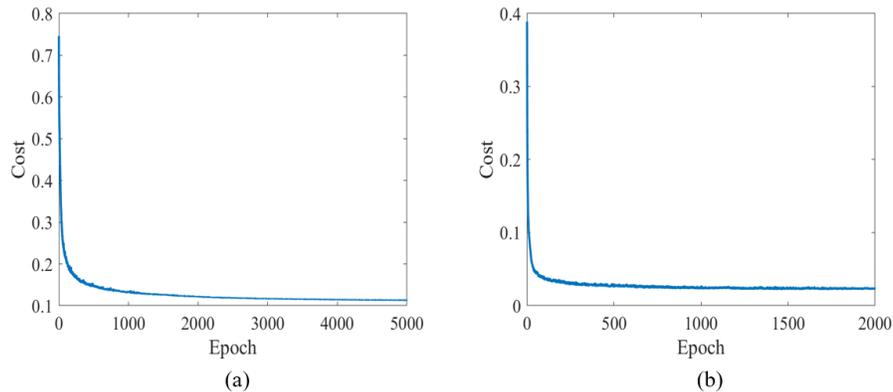

FIG. S4. Learning curve of (a) the forward modeling neural network and (b) the tandem network.